\documentclass[showpacs,aps,pre,floatfix]{revtex4}
\usepackage{graphicx} 

\begin{document}

\title{Scaling and width distributions of parity conserving interfaces}

\author{M. Arlego}

\author{M. D. Grynberg} 

\affiliation{Departamento de F\'{\i}sica, Universidad Nacional de  La Plata, 
1900 La Plata, Argentina}

\begin{abstract}
We present an alternative finite-size approach to a set of parity conserving 
interfaces involving attachment, dissociation, and detachment of extended 
objects in 1+1 dimensions. With the aid of a nonlocal construct introduced by 
Barma and Dhar in related systems [\,Phys. Rev. Lett. {\bf 73}, 2135 (1994)\,], 
we circumvent the subdiffusive dynamics and examine close-to-equilibrium
aspects of these interfaces by assembling states of much smaller, numerically 
accessible scales. As a result, roughening exponents, height correlations, and 
width distributions exhibiting universal scaling functions are evaluated for 
interfaces virtually grown out of dimers and trimers on large-scale substrates. 
Dynamic exponents are also studied by finite-size scaling of the spectrum gaps 
of evolution operators.
\end{abstract}


\maketitle

\section{Introduction}

The theory of surface growth processes has by now reached a mature status that 
allows to describe statistically a wide variety of  nonequilibrium phenomena in 
terms of universality classes of scaling regimes \cite{Krug, Odor, Henkel}. As is 
known, these latter depend strongly on the conservation laws of the underlying 
dynamics, bringing about important effects at large times. An interesting example 
of this,  introduced in the context of restricted solid-on-solid dimer growing 
interfaces \cite{Odor2, Nijs}, is the set of parity conserving (PC) processes 
\cite{Odor,Henkel}. Here, the seemingly microscopic variation of considering the 
dynamics of extended objects (i.\,e. involving more than one interface location) 
rather than that of monomers, has however far reaching implications, giving rise to 
an anomalous growth of the global roughness or interface width. In one-dimension 
(1D), where nonequilibrium roughening transitions can also take place 
\cite{Odor2}, this anomaly has been investigated in terms of even-visiting random 
walks \cite{Nijs}. In that representation the height levels of the interface are 
thought of as the visited sites of a 1D Brownian path extended on a given time 
interval, here playing the role of the substrate length $L$. The constraint to cover 
each path location an even number of times (or more generally, conserving this 
number modulo $k \ge 2$), introduces long-range temporal correlations 
\cite{Bouchaud} which causes the interface to roughen as $\sim L^{1/3}$ 
\cite{Nijs}. This is in marked contrast to the usual root mean square displacement 
of normal (diffusive) random walks characterizing the asymptotic $L^{1/2}$ width 
of a variety of interfaces grown out of monomers \cite{Krug, Odor, Henkel}, and 
typical of both 1D Edwards-Wilkinson (EW) \cite{EW} and Kardar-Parisi-Zhang 
(KPZ) \cite{KPZ} universality classes. Also, roughening anomalies were reported in 
other growth models with similar global constraints including multiparticle 
correlations \cite{Park}, self-flattening, and self-expanding surfaces \cite{Park2}.

\begin{figure}[htbp]
\includegraphics[width=1.0\textwidth]{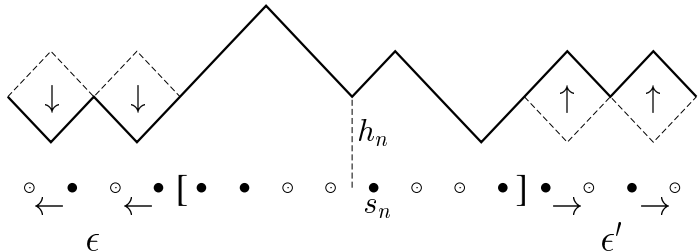}
\vskip -7.75cm
\caption{Microscopic rules of a dimer growing BCSOS interface and its 
equivalent 1D lattice gas. The dynamics of the former involves deposition  
(evaporation) of dimers with rate $\epsilon$ ($\epsilon'$) at random locations 
having at least two consecutive height minima (maxima). The corresponding 
spin-$1\over2$ ($S_n \equiv h_{n+1} - h_n$) or hard core particle dynamics 
consists, respectively, of two simultaneous left (right) particle hoppings. The 
array between brackets illustrates an irreducible string (see Sec. II\,A). Also, 
note that the identity of dimers is not necessarily maintained by subsequent
evaporations.}
\label{rules}
\end{figure}

In this work we examine further aspects of the PC processes referred to above 
focusing attention on more detailed levels of description, such as height difference 
correlation functions and width probability distributions. Interestingly and in line 
with a variety of  studies of several growth models \cite{Zia, Racz, Antal, Parisi, 
Doussal}, it will turn out that in approaching the stationary regime there is a single 
length scale, namely the usual average width, which characterizes these latter 
distributions in terms of a universal scaling function. Here we follow an alternative 
description of PC interfaces \cite{me} using a simple extension of the well known 
1D mapping between stochastic dynamics of binary lattice gases and 
body-centered solid-on-solid (BCSOS) growth processes \cite{Krug, Odor, Henkel, 
Plischke}. There, the differences of adjacent pairs of height variables $h_n$ are 
restricted to $\pm 1$, while as is shown in Fig.\,\ref{rules}, attachment and 
detachment of dimers (or $k$-mers in general), are viewed as exchanges of Ising 
spins $s_n \equiv  h_{n+1} - h_n$ on three ($2k-1$) consecutive bonds. Although 
the adsorbed particles do not diffuse explicitly neither in $k$-mer nor monomer 
form, they are allowed to rearrange throughout the interface by explicit dissociation 
of $k$-mers. This takes place under desorption attempts which may occur whether 
the $k$ targeted monomers were original adsorbing partners or not. In passing, it 
is worth mentioning that this is also a typical feature of catalytic surface reactions 
where the reconstitution of composite objects actually does matter \cite{Odor, 
Henkel, Marro}.

The simplicity of these rules is deceptive as they entail a number of conservation 
laws which grows exponentially with the substrate size. At the root of this rather
unusual partitioning of the phase space is a useful construction, namely the 
irreducible string, introduced by Barma and Dhar in closely related systems \cite
{Barma}. We shall exploit this nonlocal construct, defined later on in Sec. II\,A, 
using a simple numerical algorithm which enables an approach to the  stationary 
behavior {\it without} actually evolving the system. To that aim, we must content 
ourselves with analyzing just the situation of equal deposition-evaporation rates. 
The idea, to become clear in a moment, is to concatenate parts of steady 
configurations of initially flat but small interfaces, such that the final assembly 
also bears the global constraint of an originally flat but much {\it larger}
substrate. This approximation circumvents the problem of going through the slow 
subdiffusive dynamics characteristic of these processes \cite{Odor,Henkel,Odor2, 
Nijs, me} thus permitting a thorough sampling of width distributions in scaling 
regimes, and which otherwise would be hard to examine by standard simulations. 
Nonetheless, to probe simple features of our assembled interfaces under the 
actual dynamics, we shall also make use of those simulations in large but yet 
accessible scales. To complement this finite-size approach we also focus attention 
on dynamic exponents which, as is known \cite{Hohenberg}, can be read off from 
the spectrum gaps of evolution operators. Thus, we shall diagonalize these latter 
exactly in reachable dimensions. In line with what was mentioned before, these 
exponets will come out to be sub-diffusive and in close agreement with those 
obtained by the usual dynamic scaling of the interface width \cite{Family}.
\vskip -14.1cm

The layout of this work is organized as follows. In Sec. II we recast the master 
equation of these processes in terms of a quantum spin Hamiltonian through 
which we not only obtain dynamic exponents but also use to span the entire 
phase space of small substrates. Following Ref.\,\cite{Barma} to identify the
conservation laws of this $k$-mer dynamics and utilizing the full configuration 
lists provided by the spanning, we then put forward our algorithm to sample 
rapidly the phase space of large interfaces. In Sec. III, at first we test this sampling 
method so as to retrieve known roughening exponents as well as steady state 
correlations of smaller systems. Then we carry on with the evaluation of width 
distributions for which the onset of a universal scaling function is suggested. 
Surprisingly, this later exhibits a longer tail than that of normal random-walk-like 
interfaces \cite{Zia}. Section IV contains a concluding discussion along with brief 
remarks on open issues and possible extensions of this work.

\section{Dynamics of $k$-mer interfaces}

The stochastic dynamics of discrete Markovian systems such as those referred
to above, amounts to a generic prescription of transition probability rates $R (S
\to S') \ge 0$ between all possible  configurations $S,\,S'$ explored in time (here 
taken as being continuous). Therefore, the evolution of the probabilities $P (S,t)$ 
to  observe the system in one of these latter is controlled by a gain-loss relation
known as the master equation \cite{Kampen}, namely
\begin{equation}
\label{master}
\partial_t \,P(S,t) = \sum_{S' \ne S} \left[\: R (S'\to S)\, P(S',t)\,-
 \, R (S \to S') \,P(S,t)\:\right].
\end{equation}
Conveniently, this relation can also be rewritten in the form of a Schr\"odinger 
equation in an imaginary time, that is $\partial_t \vert P (t)\, \rangle = - H \vert P 
(t) \, \rangle$, thus permitting to derive the probability distribution $\vert P(t) \, 
\rangle \equiv \sum_S P (S,t)\, \vert\, S\, \rangle\,$ at subsequent moments from 
the action of $H$ on a given initial condition, i.e. $\vert P(t) \,\rangle = e^{- H\,t} 
\vert P(0)\,\rangle$. Here, the Liouville or evolution operator $H$ embodying the 
dynamics is defined through its matrix elements
\begin{equation}
\label{elements}
\langle\,S'\,\vert\,H \,\vert\,S\,\rangle = \cases{- R (S \to S'),\;\;\;\;\;\;\;\;\;\;\;
{\rm for}\;\;S \ne S',\cr\cr
\sum_{S'\ne S}\, R (S \to S'),\;\;{\rm for}\;\;S = S',}
\end{equation}
which, due to conservation of probability, clearly constrain all $H$-columns to 
add up to zero. Thereby it can be shown \cite{Kampen} that the steady 
state corresponds to a unique $H$-eigenmode with eigenvalue $\lambda_0 = 0$, 
whereas the relaxation time of any observable is upper bounded  by $1/{\rm 
Re}\,( \lambda_1) > 0$, with $\lambda_1$ being the first excitation level of the 
$H$-spectrum.

In our case, for what follows it is helpful to think of this evolution operator as 
being applied to a space of $1\over2$-spinors. For that, we interpret the slope 
configurations $\vert\,S\, \rangle \equiv \vert\, S_1,\dots, S_L \,\rangle$ of 
Fig.\,\ref{rules} as being already diagonal in the $z$-component, say, of Pauli 
matrices $\vec\sigma_1,\,\dots,\,\vec\sigma_L$ assigned to each slope site. By 
construction it is then clear that up to a constant $h_1$ chosen as a reference  
level, the heights of the BCSOS interface are obtained as
\begin{equation}
 h_n = h_1 + \sum_{j < n} S_j.
\end{equation} 
In particular, note that under periodic boundary conditions (PBC) the 
dynamics is consistent with a vanishing total magnetization $S^z \equiv \sum_j 
\sigma^z_j$ though as we shall see below, many further additional conservation 
laws also emerge. Introducing now the right and left $k$-mer hopping operators 
$A^+, A^-$
\begin{equation}
\label{hopping}
A^{\pm}_j = \prod_{i=1}^k \,\sigma^{\pm}_{j+2i -1}\: 
\sigma^{\mp}_{j + 2 i - 2}\,,
\end{equation}
associated respectively to the detachment and attachment processes described 
in Fig.\,\ref{rules}, and taking into account the algebra of the spin-$1 \over 2$ 
raising and lowering operators $\sigma^+_j,\,\sigma^-_j$, we can readily write 
down the operational counterpart of Eq.\,(\ref{elements}), which here reduces to
\begin{equation}
\label{H}
H =\sum_j \!\left(\,\epsilon'\,  A^+_j + \epsilon\,  A^-_j \,\right) 
\left(\, A^+_j +  A^-_j  - 1\right).
\end{equation}
Such simplicity is only apparent as the commutation algebra of the hopping  
operators complicates the analytical treatment (except for monomers and
$\epsilon = \epsilon'$, where $H$ reduces to the isotropic Heisenberg 
ferromagnet). In this latter equation evidently each of the off-diagonal terms 
provide the appropriate transition elements of the dynamics, whereas the addition 
of its diagonal parts $B^{\pm}\equiv \sum_j A^{\pm}_jA^{\mp}_j$
\begin{eqnarray}
\nonumber
B^+\!\!&=& \sum_j \,\prod_{i=1}^k\,\hat n_{j+2i -2}\,\left(1- \hat n_{j + 2 i - 1} 
\right),\;\; \hat n_j \equiv \sigma^+_j \sigma^-_j\,,\\
\label{diagonal}
B^-\!\!&=& \sum_j \,\prod_{i=1}^k\, \left( 1- \hat n_{j+2i -2}\right)\, 
\hat n_{j + 2 i - 1},
\end{eqnarray}
accounts for the number of manners $N^{\pm}_S = \langle S \vert\, B^{\pm} 
\vert\, S \rangle$ in which a given spin configuration may access to  other ones 
either by right or left jumps, i.e. $\sum_{S' \ne S} R (S \to S') = \epsilon' N^+_S + 
\epsilon \,N^-_S$ thus complying with conservation of probability. On the other 
hand combining this with the microscopic reversibility of our model, here 
expressed simply as $R (S \to S' ) = \epsilon'$ (or $\epsilon$) $\Longleftrightarrow 
R (S' \to S) = \epsilon$ (or $\epsilon'$), we then obtain
\begin{equation}
\label{identity}
\sum_{S' \ne S} \left[\: R (S' \to S) - R (S \to S')\: \right] = \left( \epsilon - 
\epsilon' \right)\,\left(\,N^+_S - N^-_S \right),
\end{equation}
from which some brief remarks about the steady state distribution now follow. 
First, note  that the monomer case is special in that for {\it PBC} (hereafter
considered throughout) this latter identity always cancels out as for $k = 1$ 
Eq.\,(\ref{diagonal}) simplifies to $B^+= B^- = {1 \over 4}\sum_j \left( 1 - 
\sigma_j^z \sigma_{j+1}^z\right)$. Therefore, comparing Eq.\,(\ref{identity})
with the right hand side of the master equation, we thus see that the monomer
steady state, either in equilibrium or not ($\epsilon \ne \epsilon'$), is consistent 
with a {\it constant} distribution \cite{Derrida}. More generally however, and 
except for the equilibrium situation, this feature does {\it not} hold for $k \ge 2$ 
because the diagonal $B^+\!, B^-$ operators of Eq.\,(\ref{diagonal}) are now 
different, so in general $N^+_S \ne N^-_S$. Since the equiprobability issue is 
essential for the sampling algorithm that follows, this  breakdown will restrict the 
numerical findings of Sec. III only to the case $\epsilon = \epsilon'$, yet being 
non-trivial for $k \ge 2$ as we shall see below.

\subsection{Irreducible strings}

Turning to conservation laws and assuming that the lattice $\Lambda =
\Lambda_1 +  \dots+ \Lambda_{2k}$ is $2k$-partite (in 1$D$ just meaning $L$ 
multiple of $2k$), the first invariant set of quantities one can readily identify 
from the composite spin exchanges of Fig.\,\ref{rules} is that of the sublattice 
magnetization differences
\begin{equation}
\label{differences}
D^z_{n,m} \equiv \sum_{j \in \Lambda_n}
\sigma^z_j - \,(-1)^{n+m}  \sum_{j \in \Lambda_m} \sigma^z_j,
\end{equation}
with $n,m = 1, \dots, 2k$. From these $2k \choose 2$ possible pairs only 
$2k-1$ of them are independent, so the number of conservation laws would grow
at most as $L^{2k-1}$. But as mentioned earlier, there is in fact a much subtler
set of constants of motion, in turn growing exponentially with the lattice size. 

\begin{figure}[htbp]
\vskip -9.3cm
\includegraphics[width=0.8\textwidth]{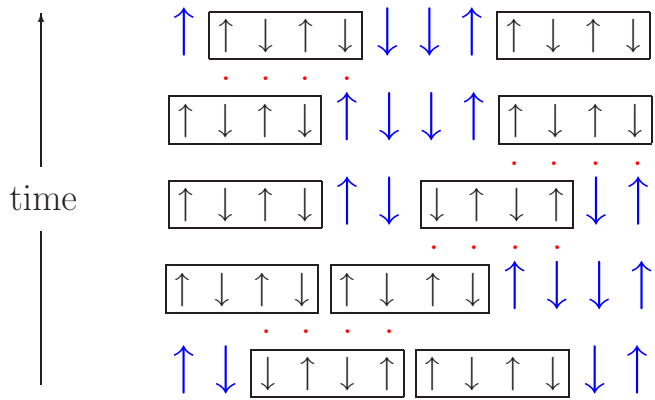}
\vskip -5cm
\caption{(Color online) Schematic random walk of irreducible characters obtained 
by the deletion process described in the text, here drawn as boxes around 
reducible groups of spins. At each step the identification of these latter is made 
from left to right.  Larger spins denote the irreducible characters whose ordering is 
left invariant by the dynamics. Dots signal the locations of updated spins.}
\label{walk}
\end{figure}

To construct that set, here we briefly survey the ideas of Ref.\,\cite{Barma}
concerning the  dynamics of deposition-evaporation of trimers reconstructing on a 
line, and which for what follows it is convenient to adapt to larger objects of even 
length, e.g. $\bullet\,\bullet\,\bullet\;\bullet \rightleftharpoons \circ\,\circ\,\circ\;
\circ$. Clearly, those processes are then isomorphic to ours via a simple
particle-hole mapping, say on even sublattices. Thus, in analogy to Ref.\,\cite
{Barma} we now define the {\it irreducible string} (IS) $I \left\{S_1, \dots\, 
S_L\right\}$ of any spin configuration as the sequence obtained by deleting all 
groups of $2k$  consecutive anti-parallel spins appearing on chosen locations, and 
then repeating recursively the procedure on the resulting shorter string until no 
further of such groups remain. As an illustration, consider for simplicity the 
following examples of $k = 2$
\begin{eqnarray}
\nonumber
I \big\{\!\uparrow \downarrow \fbox{$\!\downarrow \uparrow \downarrow 
\uparrow\!$}\downarrow \uparrow\! \big\} &=& I \big\{\!\uparrow \downarrow 
\downarrow \uparrow\fbox{$\! \downarrow\uparrow \downarrow\uparrow\!$}\,
\big\} = \big\{\! \uparrow \downarrow \downarrow \uparrow \!\big\},\\
\label{null}
I\,\Big\{\,\fbox{$\uparrow \downarrow\!$ \fbox{$\!\downarrow \uparrow 
\downarrow \uparrow\!$}\,$\uparrow \downarrow$}\,\Big\} &=& 
\big\{\,\emptyset\,\big\}, \;({\rm null\,\, string}),\\
\nonumber
I \big\{\!\uparrow \uparrow \downarrow \downarrow \uparrow  \downarrow 
\downarrow \uparrow \!\big\} &=& \uparrow \uparrow \downarrow \downarrow 
\uparrow  \downarrow \downarrow \uparrow .
\end{eqnarray}
In the first case this deletion, marked by boxes, is applied to a group of spins 
chosen either starting from the left or right. In the second instance the procedure 
is carried out recursively in two steps and no characters are left, while in the third 
example the string considered is already jammed (same irreducible block of 
Fig.\,\ref{rules}) and can no further evolve. The invariance of the irreducible 
characters  (if any) left by this process is in line with the idea that the successive 
action of the hopping operators of Eq.\,(\ref{hopping}) on a given spin 
configuration, just changes the position of those characters by multiples of $2k$ 
lattice spacings. The separations between them are mediated by substrings of 
different lengths ($\propto 2 k$), though all of these are in turn reducible to null 
strings. Thus, the interface dynamics may be thought of as a random walk of 
hard-core irreducible characters (they can not cross each other), as depicted
schematically in Fig.\,\ref{walk}. 
The positions of these walkers at a given instant of course depend on the order in 
which the reduction rule is applied, but the key issue to bear in mind here is that 
the sequence of irreducible charcaters remains unaltered throughout. In addition
and as noted in Ref.\,\cite{Barma}, two spin configurations $\vert S \rangle,\,\vert 
S' \rangle$ are connected by the dynamics $\Longleftrightarrow I \{S\}= I \{S '\}$.
Thus, the IS uniquely labels all subspaces left invariant by the  $k$-mer kinetics, 
regardless in which order the reducible groups are removed. On the other hand, it 
is clear that the number of combinations forming these irreducible sequences 
grows exponentially with the number of characters or string length ${\cal L}\le L$. 
More specifically, a straightforward analysis of a recursion relation for this length 
\cite{Barma, Robin}, shows that for large ${\cal L}$ and $k > 1$ the number of 
invariant subspaces increases as fast as $x^{\cal L}$, where $x$ is the largest root 
of $x^{2k}= 2\, x^{2k -1} - 1$.

On a more fundamental level, it would be interesting to identify the symmetries at 
the origin of these conservation laws. For instance, in the much simpler case of the 
sublattice differences of Eq.\,(\ref{differences}) the symmetries responsible for 
them just involve $\theta_1, \dots, \theta_{2k}$ rotation angles around the $z$ 
direction of each sublattice. Recalling that under those rotations $\sigma^{\pm}$ 
transform as $e^{i \pm\theta}\sigma^{\pm}$, evidently as long as $\sum_n 
(-1)^n \theta_n = 0$ is held, the hopping operators of Eq.\,(\ref{hopping}) will be 
left invariant, and so will $H$ in Eq.\,(\ref{H}). Thus, from the infinitesimal 
generators $\sum_n (-1)^n  \theta_n (\,\sum_{j \in \Lambda_n}\! \sigma^z_j\,)$ of 
these $2k -1$ independent rotations one is finally led to the conservation laws of 
Eq.\,(\ref {differences}), already obtained on more intuitive grounds. However in 
the case of the IS and the exponential proliferation of constants of motion it 
entails, the analysis appears to be much more involved. Due to the highly 
convoluted form in which the IS is obtained, unlike Eq.\,(\ref{differences}) it is 
neither clear how to construct its operational counterpart (possibly non-local), nor 
to identify the corresponding symmetries in the evolution operator. Despite that
formal insufficiency, the invariance of the IS provides an alternative computational 
tool to approach the equilibrium regime, and  which we now implement.

\vskip -0.45cm
$\phantom{hello}$
\subsection{Assembling null string states}

Notice that whenever $\rho = {\cal L}/ L$ is kept finite in the thermodynamic limit, 
the interface can not roughen at large  times \cite{me}. This is because for 
$\epsilon = \epsilon'$ the distances $\lambda$ between irreducible characters (or 
random walkers) are distributed as $\simeq \rho\,e^{-\,\rho \lambda}$ \cite
{Feller}. Thus, mean square height fluctuations along those distances (or reducible 
substrings, all with $S^z = 0$) remain bounded as $1/\rho^3$. So, hereafter we 
will focus on the null string subspace only. Besides, it is the most natural to 
consider in the context of growing interfaces, as it stems from initially flat 
conditions (plain antiferro states).

Although in equilibrium all configurations are equally weighted, yet the expectation 
value of most observables are not analytically simple to obtain because the 
ensemble of averaged states must be consistent not just with $S^z = 0$ but 
also with a 
vanishing IS. This introduces spatial correlations (absent in the monomer 
case, where $\vert \sum_j S_j \vert$ always coincides with ${\cal L}$), which 
develop slowly in the course of growth simulations. At large times however, such 
process ultimately amounts to producing a uniform distribution of null string 
states. Here we put forward an approximation to such distribution in large scale 
substrates based on the construction of small ones. The idea is to assemble pieces 
of small substrates in such a way that by applying the above deletion rules the 
whole set is reducible to the null string. 

\begin{figure}[htbp]
\vskip -0.37cm
\includegraphics[width=0.564\textwidth]{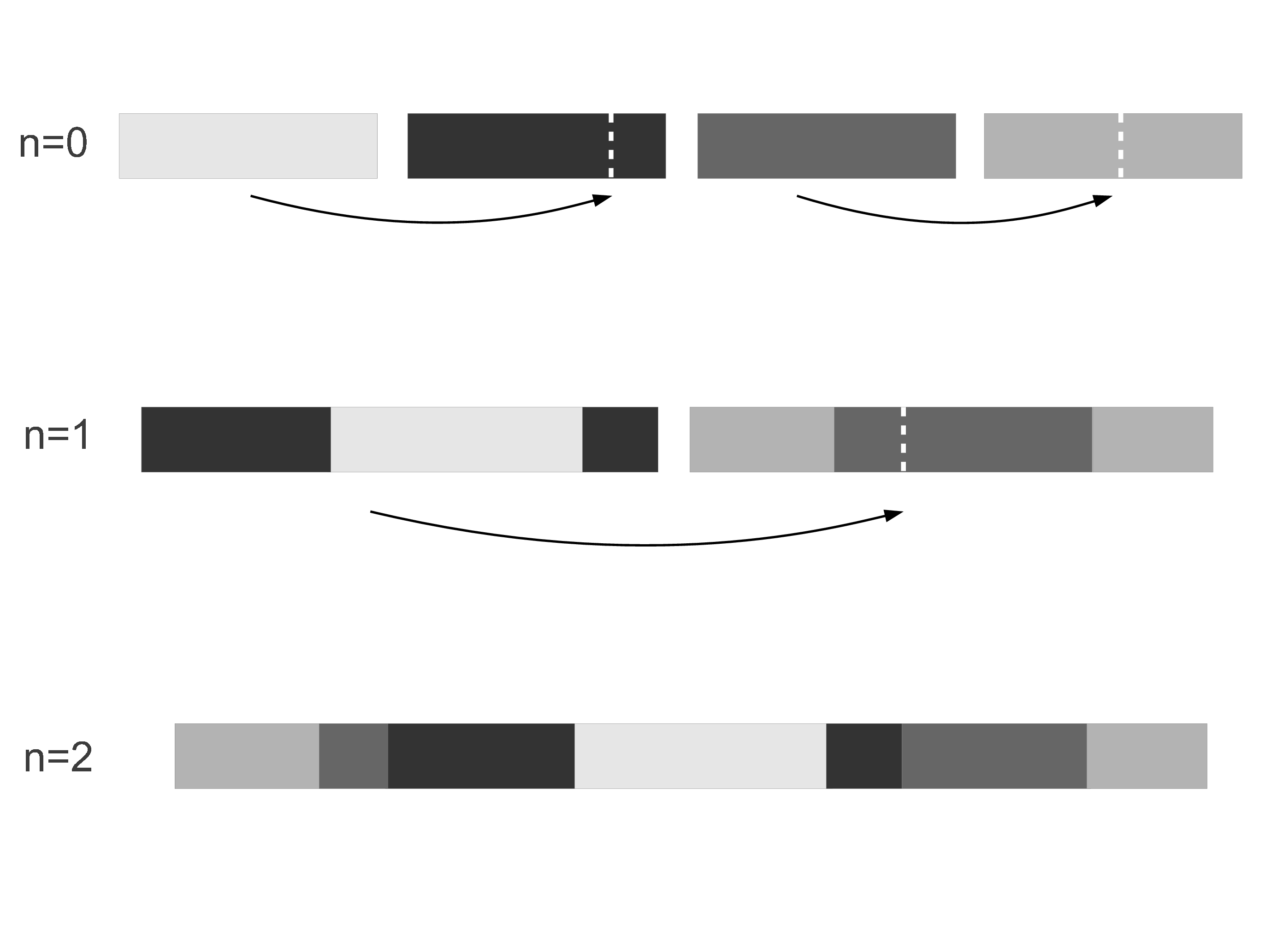}
\vskip -0.51cm
\caption{Schematic representation of the sampling algorithm after two iterations. 
The initial blocks stand for null string configurations drawn randomly from all 
possible ones constructed for an accessible lattice size $L_0$. Dashed lines denote 
random locations where an entire block is interposed. At each step the block size 
is doubled while complying with the null string constraint. Eventually, parts of the 
original blocks can also be further splitted by random intercalations.}
\label{algorithm}
\end{figure}

There are several forms to attempt this, but consider for instance a set of $2^N$ 
spin configurations (sketched as the initial blocks of Fig.\,\ref{algorithm}) with a
common length $L_0 \propto 2 k$,  and drawn randomly from a list of null string 
blocks previously prepared. This latter, in turn can be constructed from the 
repeated action of the hopping operators (\ref{hopping}) on, say, initial antiferro 
states, until exhausting the full space (typically growing exponentially with $L_0$
\cite{Barma}). At the first step, half of the drawn blocks are divided in two parts 
at random locations, while as depicted in Fig.\,\ref{algorithm}, each of the 
remaining blocks are settled between the split pairs. Thus, one is left with 
$2^{N-1}$ blocks of length $2 L_0$ all of which are evidently reducible to the null 
string [\,e.g. see Eq.\,(\ref{null})\,]. Next, the process is recursed, eventually by 
splitting further parts of the original blocks, until a single block of length $2^N 
L_0$ is obtained.  The algorithm thus generates an ensemble of substrates which 
are fully reducible by successive reductions around a central $L_0$-\,block, always 
left unsplit. Though uniform, this assembled distribution (AD) can {\it not} be 
entirely representative of the much larger substrate space, at most it just can be 
approximative. Nevertheless, as it will be tested out in Sec. III, it does reproduce 
known features of the scaling regimes in which one is ultimately interested, while 
enabling to examine there width distributions otherwise difficult to reach. Let us 
finally comment that had a simple concatenation of blocks been carried out it 
would have bounded all heights as $\vert h_j \vert \le L_0/ 2$, whereas on the 
other extreme, the use of more reduction centers neither would bring about a 
better approximation to the equilibrium regime.

\vskip -0.41cm
$\phantom{hello}$
\section{Numerical results}

Before setting out this algorithm in width distributions, we first test it against 
typical scaling aspects of growing interfaces. In studying these latter one usually 
considers  the mean square fluctuations of the average height $\bar h (t) \equiv 
{1 \over L} \sum_j h_j (t)$, which provides a measure of the global interface width 
at a given instant, that is
\begin{equation}
\label{width}
W^2 (L,t) = \frac{1}{L} \,\sum_j \left\langle \left[\, h_j (t)
\, - \,\bar h(t) \,\right]^2 \right\rangle\,.
\end{equation}
Here, the brackets denote an ensemble average over all possible evolutions of 
heights, in our case compatible with the null string imposed by flat initial 
conditions. Based on a wide range of theoretical and numerical studies, it can be 
argued that $W^2$ should scale as \cite{Krug, Odor, Henkel, Family} 
\begin{equation}
\label{scalingW}
W^2 (L,t) \simeq L^{2 \zeta}\, f ( t/L^z)\,,
\end{equation}
with a universal scaling function behaving as $f(x) \propto x^{2\zeta/z}$ for $x \ll 
1$, while approaching a constant for $x \gg 1$. Hence, for $t \ll L^z\,$ the width 
must grow as $t^{\zeta/z}$ \cite{note}, until saturating as $L^\zeta$ for times 
comparable or larger than the relaxation time $\tau$. In the above hypothesis the 
Hurst or roughening exponent $\zeta$ measures the stationary dependence of the 
interface width on the typical substrate size, while the fundamental scaling 
between length and time is given by the dynamic exponent $z$. 

\vskip -0.97cm
$\phantom{hello}$
\subsection{Scaling exponents}

When it comes to this latter it is helpful to also consider the spectral gap ${\rm 
Re}\,(\lambda_1) = 1/\tau$ of the evolution operator constructed in Sec. II, so as 
to obtain a separate evaluation (independent of $\zeta$), and which we now briefly 
touch upon. Assuming as usual the emergence of a finite-size scaling regime in 
which $\tau \propto L^z$ \cite{Hohenberg}, 
\begin{figure}[htbp]
\vskip -2.8cm
\includegraphics[width=0.526\textwidth]{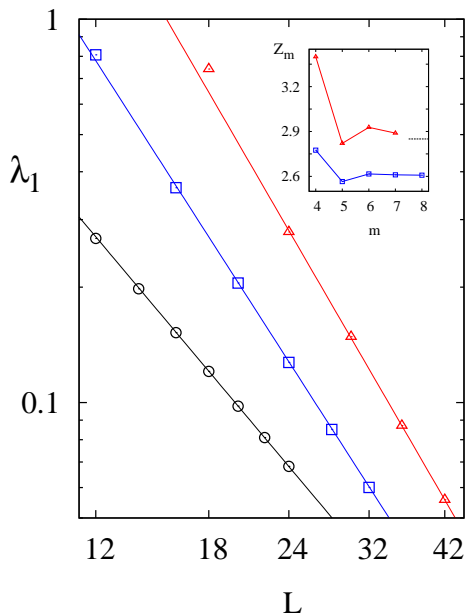}
\vskip -2.2cm
\caption{(Color online) Finite-size decay of spectral gaps of evolution operators 
[\,restricted to $\epsilon = \epsilon'$ in Eq.\,(\ref{H})\,] for dimers (squares), 
trimers (triangles), and monomers (circles, just for comparison). From top to 
bottom solid lines are fitted with slopes corresponding to dynamic exponents $z 
(k) \simeq 2.89,\, 2.61$, and 2, obtained in turn from the highest available 
approximants of Eq.\,(\ref{approx}). These are exhibited in the inset for $k= 2$ 
and 3 using several lattice sizes ($L = 2 m k)$. The horizontal line indicates the 
value of $z (3) \simeq 2.85$ arising from Fig.\,\ref{scaling}b.}
\label{gaps}
\end{figure}
we analyzed those exponents for $\epsilon = \epsilon'$ and small substrates using 
a Lanczos diagonalization \cite{Lanczos} of Eq.\,(\ref{H}) within null string 
subspaces, in turn spanned via the hopping operators (\ref{hopping}) as explained 
in Sec. II\,B. The results so obtained are displayed in Fig.\,\ref{gaps}, where it is 
seen that already modest lengths are able to yield clear finite-size trends which
evidence nonuniversal and subdiffusive slopes for dimers and trimers, i.e. $z 
\simeq$ 2.61 and 2.89 respectively. In that regard, a convergence estimation of 
these values can be made by defining the sequence of dynamic exponents or 
approximants
\begin{equation}
\label{approx}
z_m = \frac{\ln\left[ \,\lambda_1 \! \left( L_m  \right) / \lambda_1 \! \left( L_{m-1} 
\right) \,\right]} {\ln \left[ \,(m-1)/m\,\right] }\,,
\end{equation}
with $L_m  \equiv 2 k m$. As is shown in the inset of Fig.\,\ref{gaps}, the relative 
differences between our highest approximants are about 0.1\,\% for $k=2$, and 
1.3\,\% for $k=3$, which in any case are far apart from the diffusive slope ($z 
\simeq 1.99$) of the monomer case, only shown for comparison. It would be 
desirable to improve the convergence of the trimers $z$'s, but the next 
approximant ($L = 48$) requires to consider spaces of more than $2.6 \times 
10^7$ null string states which goes beyond our computing facilities. Nevertheless 
we can compare these exponents with those arising from the standard scaling 
hypothesis (\ref{scalingW}) while, more importantly, testing the validity of the 
type of scheme previously proposed.

\begin{figure}[htbp]
\vskip 0.85cm
\hskip -1.2cm
\includegraphics[width=0.436\textwidth]{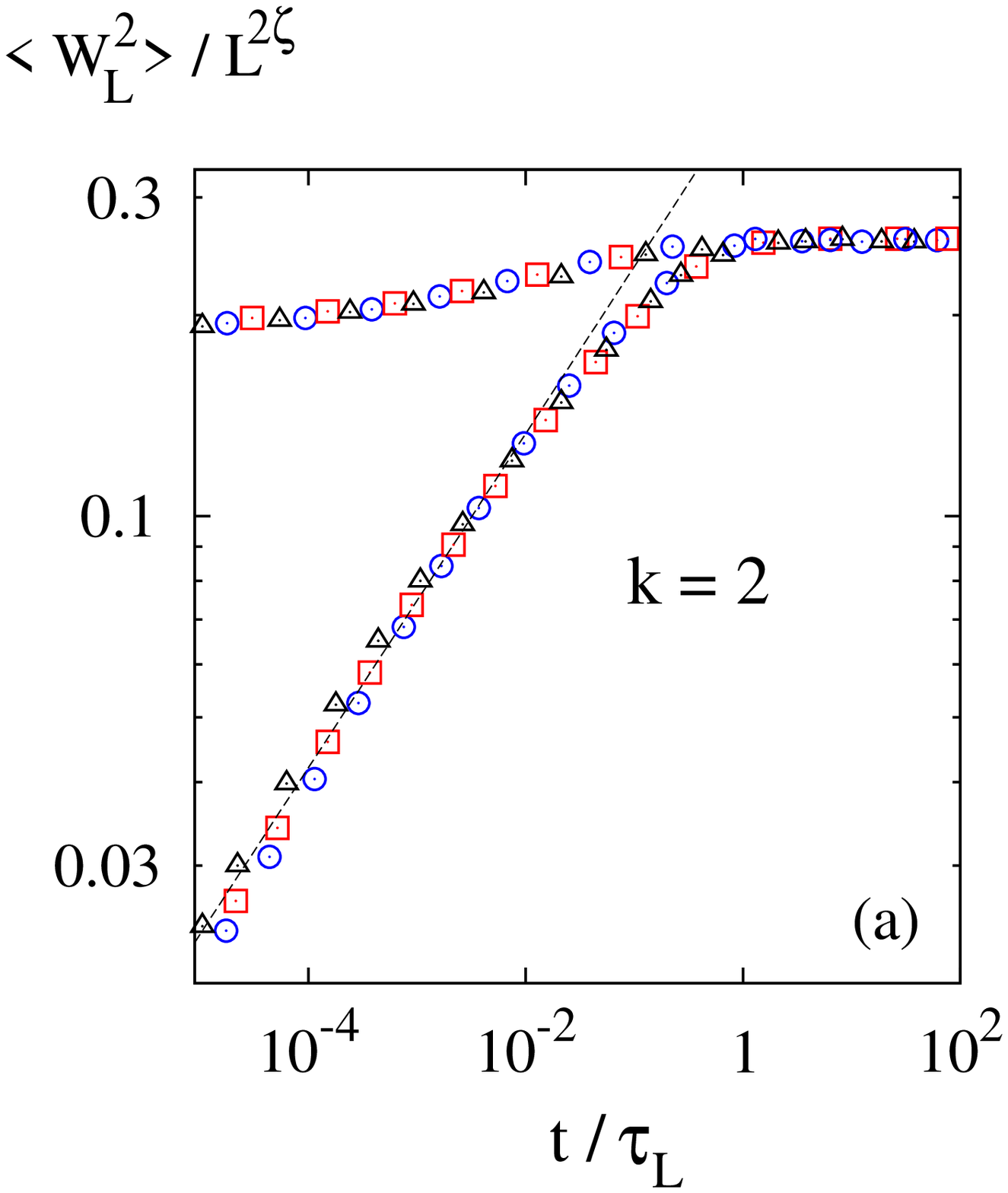}
\hskip 0.1cm
\includegraphics[width=0.436\textwidth]{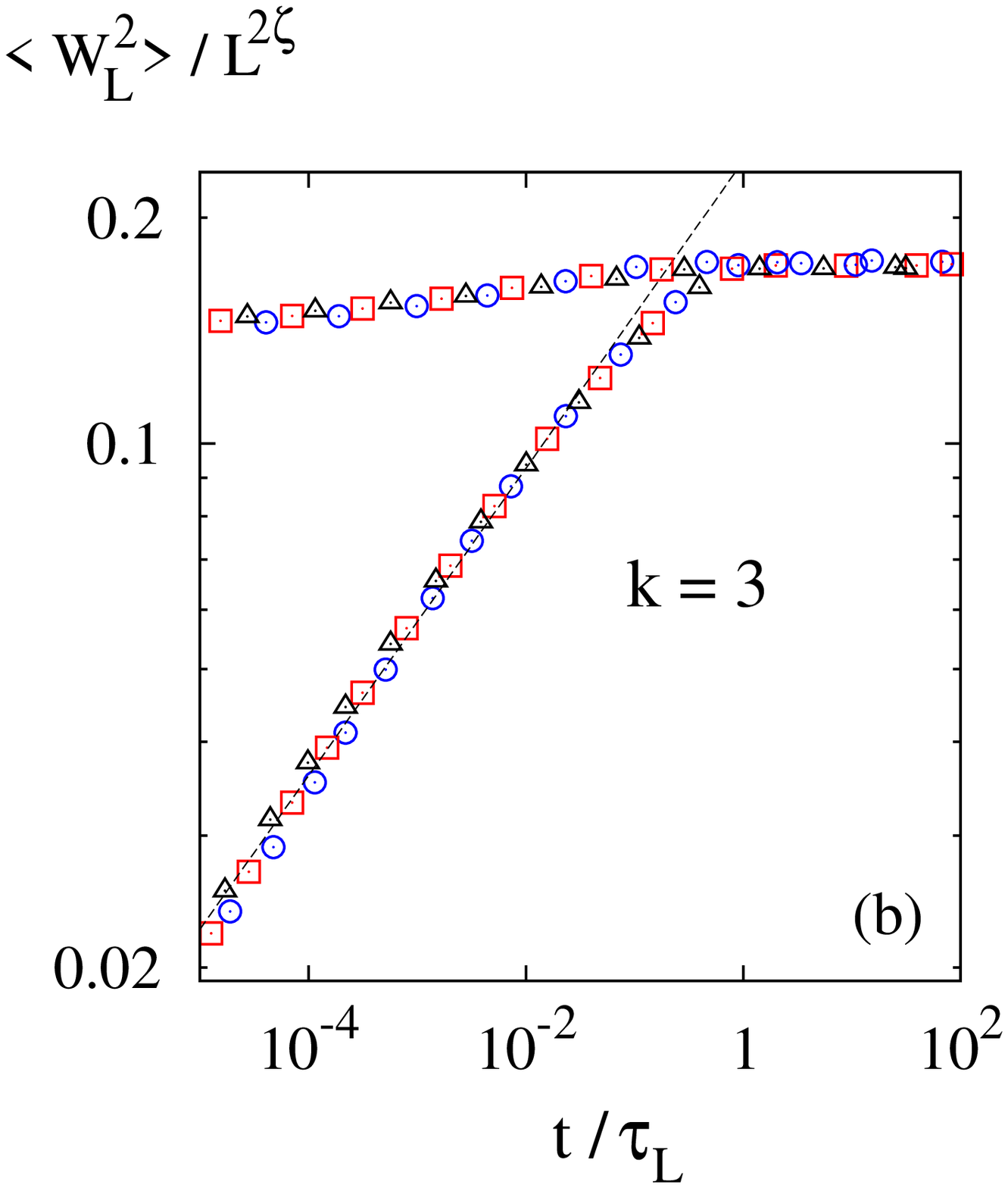}
\vskip 0.2cm
\caption{(Color online)
Dynamic scaling of interface widths using substrates of sizes $L = 2^n L_0$, for 
(a) dimers ($L_0 = 32$), and (b) trimers ($L_0 = 36$). Circles, squares, and 
triangles stand respectively for $n =$ 4, 5, and 6. The uppermost set of data was 
initially prepared from the distributions of Sec. II\,B, while the lowermost set was 
started by usual flat conditions. The scaling of data was obtained setting $\zeta 
\simeq 0.3$, and $\tau_L (k) = A_k L^{z(k)}$ with $z (2) \simeq 2.61,\, z(3) 
\simeq 2.85$, and amplitudes $A_2,\,A_3$  estimated from the reciprocal ones of 
Fig.\,\ref{gaps}. Slopes of straight lines are fitted with values $2\, \zeta /z$.}
\label{scaling}
\end{figure}

\begin{figure}[htbp]
\vskip -2.3cm
\hskip -1.1cm
\includegraphics[width=0.575\textwidth]{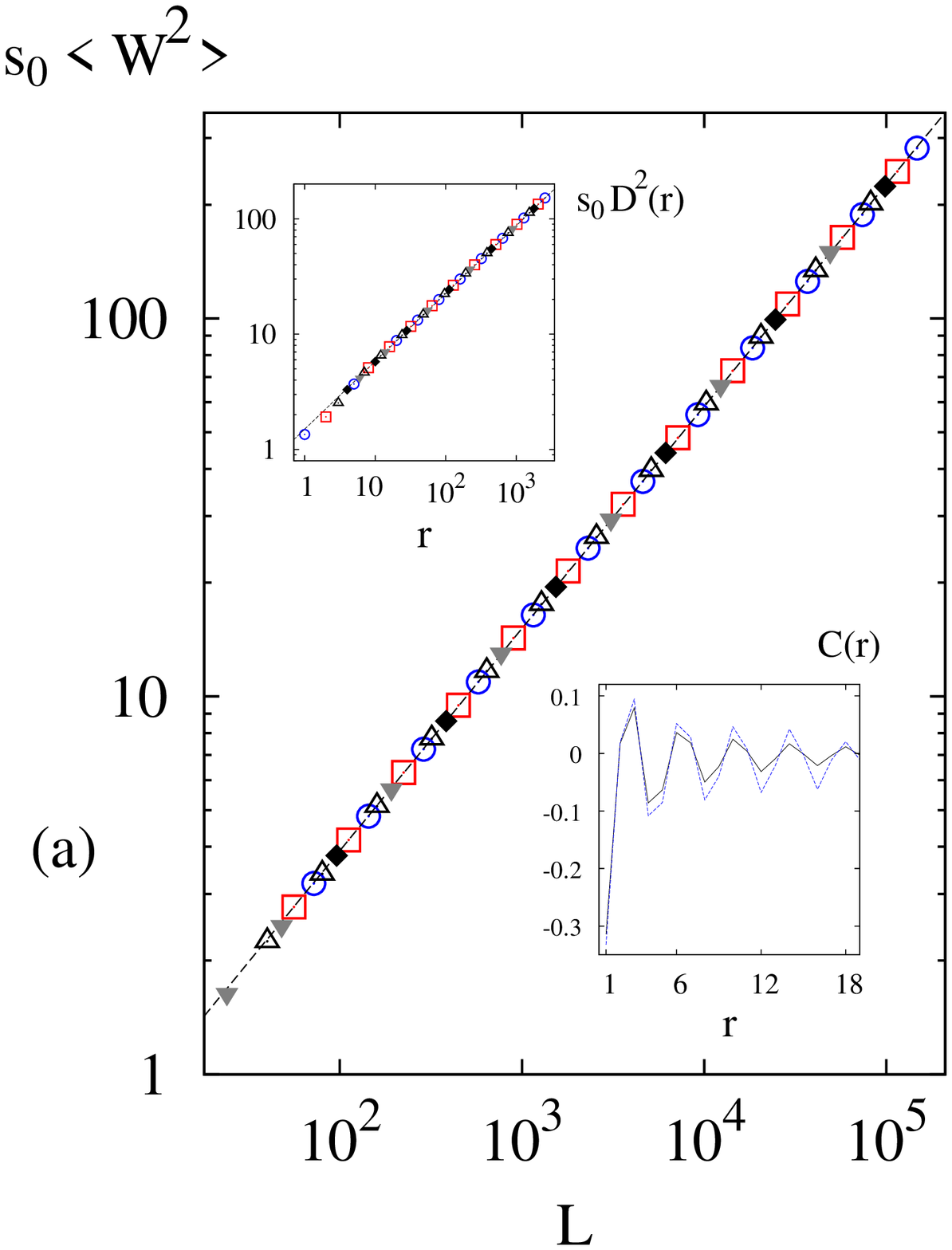}
\hskip -1.69cm
\includegraphics[width=0.575\textwidth]{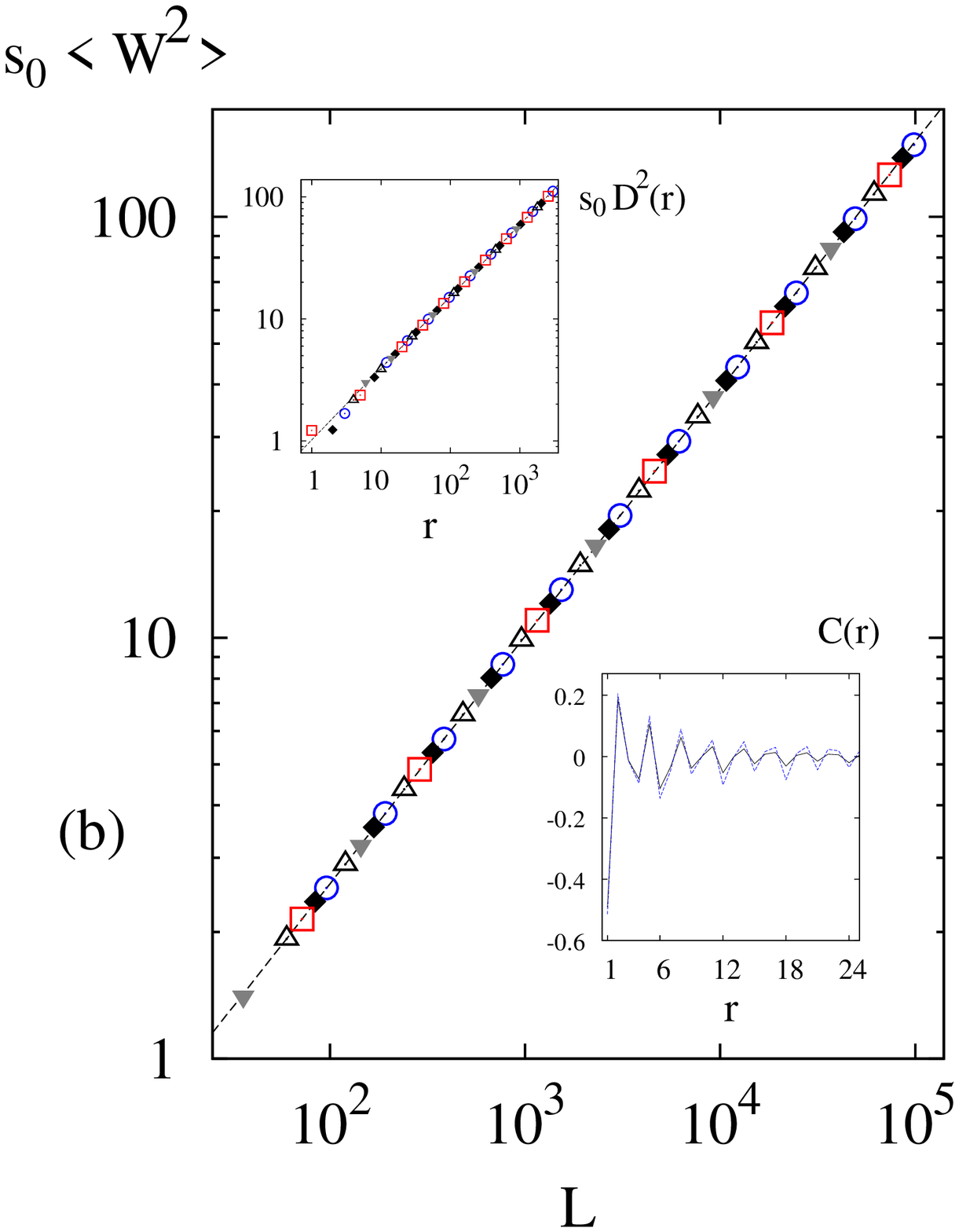}
\vskip -2.8cm
\caption{(Color online) Finite-size growth of interface width for dimers (a), and 
trimers (b) recursing the algorithm of Fig.\,\ref{algorithm}. In (a) the assembling
uses blocks of $L_0 = 12$, 20, 24, 28, and 36 sites (downwards and open upwards 
triangles, rhomboids, squares, and circles respectively), while in (b) those symbols 
stand in turn for blocks of $L_0 =  18, 30, 36, 42,$ and 48. Upper insets exhibit the 
height difference correlation functions of Eq.\,(\ref{diff-corr}) using 12 recursions. 
Each set of $L_0$-\,data was normalized by an overall scale factor $s_0$, in turn 
chosen so as to fit the saturation amplitude obtained in Fig.\,\ref{scaling}. In all 
cases the scaling $W^2 \propto L^{2 \zeta}$ and $D^2 (r) \propto r^{2 \zeta}$ 
shows up with a common roughening exponent $\zeta \simeq 0.29$. Lower insets 
display comparisons of assembled slope correlations of Eq.\,(\ref{corr}) (solid 
lines), with actual steady ones in small lattices (dashed lines).}
\label{growth}
\end{figure}

To this aim, we compared the evolutions of flat substrates with those resulting 
from the AD of Sec. II\,B.. This we do in Fig.\,\ref{scaling} where the scaled widths 
of these two different preparations are displayed for not too large sizes, so as to 
reach about $10^4$ samples in the final saturation regime (recall that $z$ is 
subdiffusive). The sets exhibit different scaling functions according to the substrate 
preparations, but in both cases the data collapse was attained upon setting a 
common roughening exponent $\zeta \simeq 0.3(1)$ (either for $k=2$ and 3), 
along with common dynamic ones $z \simeq 2.60(8)$ and $z \simeq 2.85(3)$ for 
dimers and trimers respectively. Interestingly, these two latter values happen to 
follow closely those of the approximants 
referred to above (inset of Fig.\,\ref{gaps}), while on the other hand, the early 
time widths arising from the AD already scale around a significant fraction of their 
asymptotic values, i.e. $\sim 75\,\%$ for dimers and $\sim 84\,\%$ for trimers.
This trend still improves when assembling larger $L_0$-\,blocks, namely, the scaled 
widths approach larger fractions of the saturation values observed in Fig.\,\ref
{scaling} while keeping a common roughening exponent $\zeta \simeq 0.29(3)$, 
pretty close to the value obtained above. In Fig.\,\ref{growth} this is corroborated 
for a variety of substrate sizes $2^n L_0$ ($n = 2, \dots, 12$) assembled with 
several $L_0$-\,blocks, otherwise unreachable by standard simulations. In that 
sense note that the algorithm of Sec. II\,B is not severely limited by the number of 
recursions ($n$), but rather by the large list of null string states increasing 
exponentially with $L_0$. Once these latter are evaluated, the algorithm permits to 
rapidly average over about $10^6$ samples of rather large lattices.

As we move on towards more detailed levels of description, we now consider the 
height difference correlation functions for which a similar scaling hypothesis is 
also expected to hold at distances $\vert r \vert \ll L$, that is \cite{Krug, Odor, 
Henkel, Family}
\begin{equation}
\label{diff-corr}
D_{_{\!\!L}}^2 (r,t) = \frac{1}{L} \,\sum_j \left\langle \left[\, h_{j+r} (t)\, -
 \,h_j (t) \,\right]^2 \right\rangle \simeq \vert r \vert ^{2 \zeta}\, 
g \left( t / \vert r \vert ^z\right)\,.
\end{equation}
As before, the brackets are taken as in Eq.\,(\ref{width}) and the scaling function
$g(x)$ behaves analogously to that of Eq.\,(\ref{scalingW}). From this, one infers 
that a time $t \sim \vert r \vert^z$ is required to fully develop the interface 
roughness across a given distance $\vert r \vert$. Thus, to check out whether our
AD can already be associated to late stages of growth, we measured these height 
correlations using over three distance decades for several $L_0$-\,substrates. The 
results are shown in the upper insets of Fig.\,\ref{growth} where it is observed 
that in all cases the previous roughening exponent $\zeta \simeq 0.29(3)$ is 
recovered. For displaying purposes we rescaled these data with the same 
$s_0$-\,amplitudes ($1.3 \alt s_0 \alt 1.5$) used in the main panels, which is no 
coincidence, as it would be expected on the basis of the identity 
$\lim_{_{_{\!\!\!\!\!\!\!\!\!\!\!r \to \infty}}} 
\lim_{_{_{\!\!\!\!\!\!\!\!\!\!\!L \to \infty}}}\!\!D_{_{\!\!L}}^2 (r) = 
\lim_{_{_{\!\!\!\!\!\!\!\!\!\!\!L \to \infty}}}\!\!2\, W_{_{\!\!L}}^2$.

Before continuing we pause to comment on the differences appearing between the 
exponents of the dimer dynamics and those in the even-visiting random walks 
(EVRW) analyzed in Ref.\,\cite{Nijs}. Our $\zeta$ value should {\it not} be 
regarded as a mere numerical deviation from the $1/3$ exponent conjectured in 
\cite{Nijs}. At the origin of this departure is the exponential proliferation of 
irreducible strings appearing in the dimer dynamics which ultimately impose 
tighther restrictions than those already occurring in EVRW.  Although both 
dynamics  share the topological constraint caused by the mod 2 conservation of the 
number of particles at every height level, note that the BCSOS version of EVRW 
\cite{Nijs} mixes up all the many-sector decomposition discussed in Section II\,A. 
For instance, besides the dimer dynamics Ref.\,\cite{Park} also considers the full 
restoration of ergodicity by deposition-evaporation of two particles at two randomly 
chosen columns with  equal heights. These need not to be contiguous (as in the 
dimer dynamics), nor necessarily share the same terrace. That introduces a 
genuine one-to-one correspondence with the ensemble generated by the EVRW 
dynamics, so the estimate of $\zeta \sim 0.33$ found there in such conditions is 
then in line with the theory of Ref.\,\cite{Nijs}. Similarly, that latter work also 
investigates the effect of adding monomer diffusion within terraces. That produces 
another estimation which yields $\zeta \sim 0.31$. But further to that difference, 
note that the explicit addition of monomer diffusion partially relaxes the broken 
ergodicity of the original dimer dynamics. Hence, the above deviations of $\zeta$ 
from $1/3$ should not be ascribed merely to statistical errors but mainly to the 
change of conservation laws. In fact, when restoring full ergodicity Ref.\,\cite{Park} 
yields a KPZ type exponent $z \sim 1.5$ which is far apart from the subdiffusive $z 
\sim 2.6$ obtained in the original dimer dynamics of both Refs.\,\cite{Nijs} and 
\cite{Park} as well as in this Section.

Turning to smaller scales and to further probe the AD, we finally compare the 
exact slope or spin correlations
\begin{equation}
\label{corr}
C (r) = {1 \over L} \sum_j \left\langle \, S_{j+r}\,S_j\, \right\rangle,
\end{equation}
evaluated in the uniform distributions of our largest available blocks (36 heights 
for dimers and 48 for trimers), with those estimated in substrates assembled with 
{\it smaller} $L_0$'s. This is illustrated in the lower panels of Fig.\,\ref{growth} 
where it is corroborated that in both cases these functions closely approximate 
each other. As mentioned earlier on, we thus see that even though the averaged 
distributions are uniform, the null string constraint enforces non-trivial 
correlations which otherwise would not appear by the sole restriction of $S^z = 0$ 
(as in the case of monomers, where all even correlators $\left\langle \,S_{j_1} 
\dots \,S_{j_{2n}} \right\rangle$ vanish identically as $L^{-n}$). Noting that 
either Eq.\,(\ref{width}) or (\ref{diff-corr}) can also be expressed in terms of 
$C(r)$, it follows that for $k \ge 2$ these pair correlations are the ultimately 
responsible for the anomalous roughening of PC interfaces, as opposed to the case 
$k=1$ where these pairs ($\propto L^{-1}$) have no effect.

\subsection{The width distribution}

It is reassuring that both roughening exponents and correlations of Fig.\,\ref
{growth}, in conjunction with the near-saturated scaling regimes of Fig.\,\ref 
{scaling}, all suggest  strongly that the construction of Section II\,B is sampling 
close-to-equilibrium states. Thus, we carry on and further exploit that construction 
to compute the width distributions of PC interfaces.

Since the dynamic scaling hypothesis (\ref{scalingW}) involves in fact an integral 
over all interface modes, it may well occur that corrections to scaling are needed. In 
our case, this is particularly noticeable at early stages of evolution where the data 
collapse in the lower sets of Figs.\,\ref{scaling}a and \ref{scaling}b is not so 
evident. In that sense, a variety of theoretical and numerical studies \cite{Zia,Racz, 
Antal,Parisi,Doussal} have suggested an alternative characterization of interfaces 
in terms of the full probability distribution $P\left( w^2 \right)$ of its particular 
random width realizations $w^2$. So long as their average  $\langle\, W^2 \, 
\rangle$ diverges in the thermodynamic limit, i.e. $\zeta >0$, the relevance of such 
distribution relies in that for large substrates it scales as \cite{Zia, Racz, Antal, 
Parisi, Doussal}
\begin{equation}
\label{scalingP}
P_{\!\!_L}\left( w^2 \right) \simeq {1 \over \langle\, W_{\!\!_L}^2 \,\rangle}\,
\Phi \left({w^2 \over \langle\, W_{\!\!_L}^2 \,\rangle}  \right)\,,
\end{equation}
where the scaling function $\Phi (x)$ is a {\it universal} characteristic of the 
interface fluctuations. Like $\zeta$ however, at large times this function can not 
point out dynamic aspects of universality classes. For example, since in 1D the 
steady sates grown out of $k=1$ are equiprobable and uncorrelated (recall Sec. II), 
$\Phi (x)$ is the same for both EW ($z = 2,\,\zeta = {1 \over 2}$) 
and KPZ ($z = {3 \over 2},\, \zeta = {1 \over 2}$) universality 
classes \cite{Zia}. But in view of the role of null strings for $k \ge 2$,\, $\Phi (x)$ 
should be able to distinguish clearly the growth of PC interfaces from that of 
monomer ones. 

To that aim, note that the basic problem of sampling stationary $w^2$ under very 
slow relaxation ($z > 2$) inevitably arising in standard simulations, is to a large 
\begin{figure}[htbp]
\vskip -2cm
\includegraphics[width=0.598\textwidth]{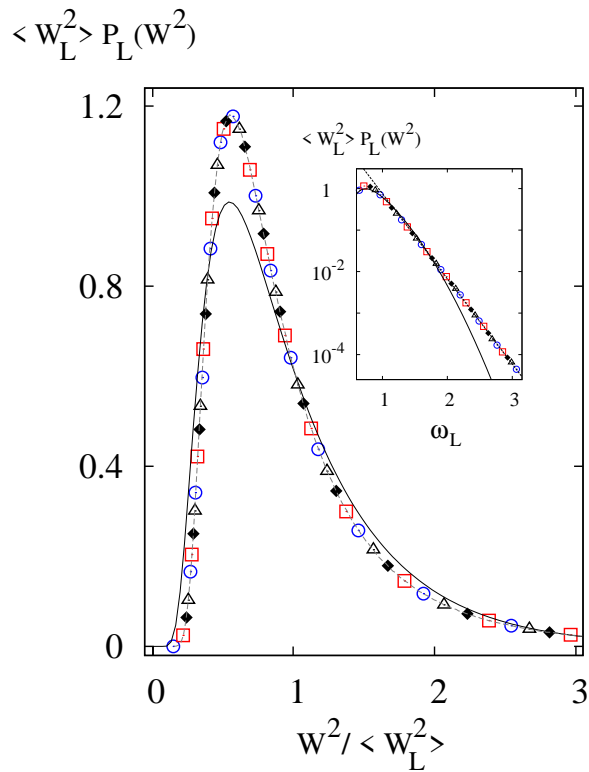}
\vskip -2.8cm
\caption{(Color online) Scaling of width distributions close to equilibrium regimes 
of dimer and trimer interfaces for various lattice sizes ($2^n L_0$). Squares and 
triangles stand for strings of $k = 2$ assembled with $L_0= 36$ and 24 using 
$n = 10$ and 8 respectively. In turn, circles and rhomboids denote $k = 3$ strings 
stemming from blocks of $L_0 = 48$ and 30, also recursed 10 and 8 times in each 
case.  For comparison, the solid line refers to the $k = 1$ distribution of Ref.\,\cite
{Zia}, whereas the dashed one is just a guide to the eye. The inset evidences a 
longer tail for $k \ge 2$, clearly decaying exponentially with $\omega_{\!_L} \equiv 
W / \sqrt{\,<\!\!W_{\!\!L}^2\!\!>}\,$.}
\label{distribution}
\end{figure}
extent bypassed by the close-to-equilibrium distributions of Sec. II\,B. In turn, these 
also enable to sample sufficiently large lattices as otherwise there would be a 
rather small number of possible $w^2$ values coarse-graining the histogram of 
$P_{\!\!_L}\!\left(w^2 \right)$.  Thus, using substrates in the range of 6.000 to 
50.000 heights, in Fig.\,\ref{distribution} we display the scaled distributions 
obtained after dividing the data into $\simeq 10^3$ intervals (not all shown), and 
averaging over about $5 \times 10^7$ samples. This rather extended sampling, 
facilitated through the algorithm of Sec. II\,B, was necessary to capture the 
statistics of the rare events on the tail of the distribution given in the figure inset. 
As expected, in all cases the data collapse into a single scaling function easily 
discernible from the exact monomer result \cite{Zia}. (The collapse is better 
demonstrated within the scales of the inset where probabilities are much smaller). 
Furthermore, $\Phi (x)$ appears to be the same for dimers and trimers, each 
stemming from dissimile types of assembled null strings, thus suggesting a 
universal function for equilibrium PC interfaces. In contrast to the scaling 
hypothesis (\ref{scalingW}), here note that there are no parameters to collapse the 
$\Phi$'s, and no scaling properties of $\langle\,W_{\!\!_L}^2\,\rangle$ are used or 
assumed. The only approximation is the finite size of the systems investigated. In 
that regrad, other large substrates assembled with different $L_0$'s and $n$'s (not 
displayed to avoid overcrowding) yielded the same, numerically indistinguishable, 
functions.

Just as the monomer case, we see that the length scale $L^{\zeta}$ not only 
characterizes the macroscopic level of the interface roughness [\,Eq.\,(\ref 
{scalingW}) and Fig.\,\ref{growth}\,], but also emerges as the natural length of the 
whole width distribution. Since $\zeta < 1/2$, one would intuitively presume that 
height fluctuations in PC interfaces are smaller than those in monomer ones. Thus, 
on approaching the above length scale one would expect $\Phi (x)$ to become 
more peaked and narrower than the monomer $\Phi$, something which in fact  
occurs to some extent. However, on the same basis one would also expect the PC 
distribution to decay faster for large width realizations. Surprisingly, however it 
turns out to be the other way around. This is illustrated in the inset of Fig.\,\ref
{distribution} where the semilog plot strongly suggests an exponential decay in 
the scaled variable $W/\sqrt{\langle\,W^2\rangle\,}$ rather than its squared, as 
occurring in the exact solution of $k=1$ \cite{Zia}. More specifically, the tails of 
these two distributions behave as
\begin{equation}
\label{tails}
\Phi (x > 1.5) \simeq \cases{ {\pi^2 \over 3}\, \exp\left( - \,{\pi^2  
\over 6}  x \right),\;\, {\rm for}\;\; k = 1\,,\cr\cr
a \,\exp\big(- b \,\sqrt {x\,}\; \big),\;\, {\rm for}\;\; k = 2, 3 \,,}
\end{equation}
with fitting parameters $a \simeq 85 \pm 2$ and $b \simeq 4.7 \pm 0.1$. Hence, 
we conclude that even though the average roughness of finite PC interfaces is 
significantly smaller ($\zeta \simeq 0.29$) their fluctuations can eventually explore 
larger widths.

\section{Summary and discussion}

To summarize, we have presented an alternative approach to 1D parity conserving
interfaces close to their equilibrium regimes ($\epsilon = \epsilon'$). The 
notion of irreducible string \cite{Barma}, which partitions the dynamics into very 
many disjoint sectors of the configuration space, played an instrumental role in the 
implementation of the assembling algorithm put forth in Sec. II\,B (Fig.\ \ref 
{algorithm}). This latter was shown to provide a fair sampling of the almost 
saturated state, in turn exhibiting a scaling regime (Fig.\,\ref{scaling}) controlled 
by the very same dynamic and  roughening exponents obtained through the 
dynamic scaling hypothesis (\ref{scalingW}) \cite{Family}. Without explicitly 
evolving the system in time, the assembled distribution also reproduced that latter 
exponent both at the macro scale of the average interface width (main panels of 
Fig.\,\ref{growth}), as well as at the micro level of the height difference 
correlations of Eq.\,(\ref{diff-corr}) (upper insets). The value of $\zeta \simeq 
0.29(3)$ so obtained is also in excellent agreement with those resulting from  
simulations in previous studies \cite{me} and with restricted solid-on-solid 
versions of these interfaces \cite{Odor2, Nijs}. 

As for dynamic exponents, we diagonalized the evolution operator (\ref{H}) 
within the accessible null string spaces already stored in assembling the above 
distributions, so as to analyze the size dependence of its spectral gap (Fig.\,\ref
{gaps}). This provided an estimation of dynamic exponents via a sequence of 
finite-size approximants (inset of Fig.\,\ref{gaps}) nearing subdiffusive but 
nonuniversal values, i.e. $z \simeq 2.61$ and 2.89 for dimers and trimers 
respectively, both in reasonable agreement with the exponents resulting from 
the scaling hypothesis (\ref{scalingW}).

Apart from avoiding the slow subdiffusive dynamics, perhaps the most interesting 
aspect of our assembling approach is that it also allows for a rich statistical analysis 
of the full width distribution of large scales. All assembled sizes yielded a single 
universal scaling function for both dimers and trimers (Fig.\,\ref{distribution}), in 
turn quite distinct from that of normal random-walk or monomer interfaces \cite
{Zia}. This contributes to the list of already known scaling functions \cite{Zia,Racz, 
Antal,Parisi,Doussal} which concurrently with roughening exponents may be used 
to identify static universality classes of growth processes. The only length scale 
spontaneously emerging in those functions is the average interface width $\propto 
L^{\zeta}$, which in the PC class results conspicuously smaller than in other 
classes. Curiously, however, height fluctuations in this former turn out to build up 
in such a way that tails of width distributions decay much slower than those in 
monomer interfaces (stretched exponential of Eq.\,(\ref{tails}) and inset of 
Fig.\,\ref{distribution}).

Turning back to irreducible strings, note that in equilibrium the only one in which 
fluctuations diverge even in the thermodynamic limit is just the null string. All 
other ones containing a finite density $\rho = {\cal L}/L$ of irreducible characters 
might be considered as non-critical strings. In that regard, Fig.\,\ref{walk} is
helpful to understand these latter as random sequences of null substrings of length 
$\lambda$ (distributed as $\rho\,e^{- \rho\, \lambda}$ \cite{Feller}), through 
which mean square height fluctuations can not but to remain bounded as $\sim 
\int_0^{\infty} \!\lambda^2 \,\rho\,e^{- \rho\, \lambda} d \lambda = 2/\rho^3$. 
However note that as soon as $\epsilon \ne \epsilon'$, stationary probabilities 
immediately become nonuniform [\,recall discussion below Eq.\,(\ref{identity})\,] 
and this simple picture no longer holds. Notwithstanding that the nonequilibrium 
dynamics is still partitioned by the same strings, it remains to figure out whether 
the current assembling approach, either for null or finite strings, could be extended 
to incorporate those nonuniform measures. Because of these latter, nonequilibrium 
width distributions need no longer be related to the scaling function obtained here. 
Finally, in $d > 1$ where there is no analogue of irreducible string, all these issues 
either in equilibrium or not remain quite open.

\vskip -0.74cm
$\phantom{hello}$
\section*{Acknowledgments}

It is a pleasure to thank helpful discussions with F. A. Schaposnik. The authors 
acknowledge support of CONICET and ANPCyT, Argentina, under Grants PIP 1691 
and PICT 1426.



\begin{thebibliography}{99}
\bibitem{note} However for the 1D PC interfaces being considered, the specific
form of this scaling hypothesis holds only under detailed balance conditions. 
Otherwise, the width either exhibits a logarithmic growth or is kept bounded; see 
Refs.\,\cite{Odor2, Nijs, me}.
\end{thebibliography}
\end{document}